\documentclass{iopart}   
\usepackage{iopams}  
\usepackage{epsf}  
\usepackage{times}  
 
\eqnobysec

\begin{document}

\jl{1}

\paper{A General Integral Identity }

\author {ML Glasser}

\address{\ Department of Physics, Clarkson University, Potsdam\\ NY 13699-5820, USA}
\address{\ Departamento de F\'isica Te\'orica, At\'omica y \'Optica,
Universidad de Valladolid, Valladolid 470071, Spain}

\begin{abstract}
The identity
$$\int_0^{\pi/2}d\phi\int_0^{\pi/2}d\theta\sin\phi F(x\sin\phi\sin\theta)=\frac{\pi}{2}\int_0^1F(xt)dt$$
where $F$ is any function, is derived. Several extensions are given and a few examples of physical interest are described.
\end{abstract}

\vskip .2in
\begin{quote}
6accdae13eff7i3l9n4o4qrr4s8t12ux

\vskip .1in
Isaac Newton, 1676
\end{quote}
\vskip .3in\noindent


\section{Derivation}

We begin by noting two formulas from the familiar reference[1]: The Bessel transform (6.567(13))($x>0)$

$$\int_0^1\frac{dt\; t}{\sqrt{1-t^2}}I_0(xt)=\frac{\sinh(x)}{x}\eqno(1)$$
and the definition of the Struve function
$${\bf L}_0(z)=\sum_{m=0}^{\infty}\frac{(z/2)^{2m+1}}{\Gamma^2(m+3/2)}.\eqno(2)$$
From (2) with $z=xt$, substituting $t^2=u$, integrating term-by-term and using Euler's beta integral we find
$$\int_0^1\frac{dt\; t}{\sqrt{1-t^2}}{\bf L}_0(xt)=\frac{\sqrt{\pi}}{2}\sum_{m=0}^{\infty}\frac{(x/2)^{2m+1}}{\Gamma(m+3/2)\Gamma(m+2)}=\frac{\cosh(x)-1}{x}.\eqno(3)$$
Consequently,
$$\int_0^1\frac{dt\; t}{\sqrt{1-t^2}}[I_0(xt)-{\bf L}_0(xt)]=\frac{1-e^{-x}}{x}=\int_0^1e^{-xt}dt.\eqno(4)$$
However[2],
$$\frac{\pi}{2}[I_0(z)-{\bf L}_0(z)]=\int_0^1\frac{dt}{\sqrt{1-t^2}}e^{-zt}.\eqno(5)$$
Therefore,
$$\int_0^1\frac{u\; du}{\sqrt{1-u^2}}\int_0^1\frac{dt}{\sqrt{1-t^2}}e^{-xut}=\frac{\pi}{2}\int_0^1e^{-xt}dt\eqno(6)$$
which can be written
$$\int_0^{\pi/2}d\phi\int_0^{\pi/2}d\theta\sin\phi e^{-x\sin\phi\sin\theta}=\frac{\pi}{2x}\int_0^xe^{-t}dt.\eqno(7)$$

Next, let $F$ belong to the class of functions which are Laplace transforms, i.e. for some real $f$
$$F(s)=\int_0^{\infty}e^{-st}f(t)dt\eqno(8)$$
Now, in (7) replace $x$ by $xu$, multiply both sides by such an $f(u)$ and integrate both sides with respect to $u$ over $[0,\infty]$, with the further restriction (on $f$)  that the order of integration can be freely interchanged. The result is
$$\int_0^{\pi/2}d\phi\int_0^{\pi/2}d\theta\sin\phi F(x\sin\phi\sin\theta)-\frac{\pi}{2x}\int_0^xF(t)dt=0.\eqno(9)$$
The left hand side of (9) is a linear functional on an ideal of the class of all real integrable functions. The Hahn-Banach lemma[3] assures that (9) can be extended, at least to the class of all piecewise continuous real-valued functions on the positive real line.
\vskip .1in

Since $x$ is a free parameter, let us replace $x$ by $x\sin\beta$ and integrate over $\beta$ to get
$$\int_0^{\pi/2}d\beta\int_0^{\pi/2}d\phi\int_0^{\pi/2}d\theta\sin\phi F[x\sin\beta\sin\phi\sin\theta]=\frac{\pi}{2x}\int_0^{\pi/2}\frac{d\beta}{\sin\beta}\int_0^{x\sin\beta}dt F[t].\eqno(10)$$
By integrating the right hand side of (10) by parts, followed by the substitution $u=\cos\beta$ we have the intriguing identity
$$\int_0^1\int_0^1\int_0^1\frac{ududvdw}{\sqrt{(1-u^2)(1-v^2)(1-w^2)}}F(xuvw)=$$
$$\frac{\pi}{4}\int_0^1du\ln\left(\frac{1+\sqrt{1-u^2}}{1-\sqrt{1-u^2}}\right)F(xu).\eqno(11)$$
Clearly, this process can be repeated to obtain a reduction formula for an $n-$fold multiple integral, $n=4,5,6,\dots$, to at most an $(n-2)$-fold integral. 

\vskip .1in
It  seemed unlikely that such a general and useful identity as (9) has no precedents in the classical mathematical literature; a bit of searching showed that  such a connection does exist. Let us set $F(x)=f'(x)$. and note that the second term on the left hand side of (9) is formally
$[f(x)-f(0)]/x$.  Next write $h(x)=f(0)+x\int_0^{\pi/2}d\phi\sin\phi f'(x\sin\phi)$, so (9) can be expressed $f(x)=(2/\pi)\int_0^{\pi/2}h(x\sin\theta)d\theta$, which is precisely Schl\"omilch's integral equation[4]. Thus it is possible that Schl\"omilch, or Abel, since his eponymous equation is equivalent to Schl\"omilch's, was aware of some form of (9).

It is reasonable, therefore, to seek a more formal derivation of (9) that does not rely on manipulating specific functions. The following argument places it  in principle among a  family of such identities and displays its elementary character.  Consider the double integral
$$S=\int_0^1dx\int_0^1dy\frac{f(x+y)F(xy)}{\sqrt{(1-x^2)(1-y^2)}}.\eqno(12)$$
where $f$ and $F$ are arbitrary. The change of variables
$$x=\frac{1}{2}[u+\sqrt{u^2-4v}]$$
$$y=\frac{1}{2}[u-\sqrt{u^2-4v}]\eqno(13)$$
having Jacobian $(u^2-4v)^{-1/2}$ leads to
$$S=2\int_0^1dvF(v)\int_{2\sqrt{v}}^{v+1}dv\frac{f(u)}{\sqrt{(u^2-4v)[(1+v)^2-u^2]}}.\eqno(14)$$
Now let $f(x)=x$. The $u-$integral is $\pi/2$ yielding (9) after a trigonometric substitution. Every choice of $f$ gives a possibly new integral identity. For example take $f(x)=1$. Since
$$\int_{2\sqrt{v}}^{1+v}\frac{du}{\sqrt{(u^2-4v)([(1+v)^2-u^2]}}=\frac{1}{v+1}{\bf K}\left(\frac{1-v}{1+v}\right),\eqno(15)$$
 after an elementary change of integration variable, (12) becomes
$$\int_0^1\frac{dt}{1+t}{\bf K}(t)F\left(\frac{1-t}{1+t}\right)=\frac{1}{2}\int_0^{\pi/2}d\theta\int_0^{\pi/2}d\phi F(\sin\theta\sin\phi).\eqno(16)$$
As an application of (16), consider the family of integrals
$$K_n=\int_0^1\frac{{\bf K}(k)}{(1+k)^n}dk\eqno(17)$$
of which only the members $n=0,1$ appear to be well- known [1,7]. First, with $F(x)=1$ we get immediately $K_1=\pi^2/8$. Now, by setting $F(x)=x^n$, writing $1-k=2-(1+k)$ and employing the binomial series, we find the recursion relation
$$K_{n+1}=\frac{\pi}{2^{n+3}}\left[\frac{\Gamma^2\left(\frac{n+1}{2}\right)}{\Gamma^2\left(\frac{n+2}{2}\right)}-(-1)^n\pi\right]-\sum_{k=1}^{n-1}(-1)^k\left(\begin{array}{c}
n\\
k
\end{array}\right)2^{-k}K_{n+1-k}\eqno(18)$$
$$n=1,2,3,\dots$$

An interesting formula results from selecting $F$ in such a way as to cancel the elliptic integral in (16):
$$\int_0^{\pi/2}d\theta\int_0^{\pi/2}d\phi\frac{1}{{\bf K}\left(\frac{1-\sin\theta\sin\phi}{1+\sin\theta\sin\phi}\right)}=2\ln(2).\eqno(19)$$

\vskip .2in
\centerline{\bf Other Examples}\vskip .1in

\vskip .1in

In a recent study of Feynman diagrams in two-dimensional quantum field theories[5] and related work[6] attention was drawn to various moments of powers of the complete elliptic integral of the first kind ${\bf  K}(k)$. A number of these were evaluated and the values of several others were conjectured.  The form of (9) strongly suggests that it may prove useful in this connection. For example, we have[7]
$$\int_0^{\pi/2}{\bf K}(a\sin\theta)d\theta={\bf K}^2\left(\sqrt{\frac{1-\sqrt{1-a^2}}{2}}\right)\eqno(20)$$
$$\int_0^1{\bf K}(xt)dt=\frac{\pi}{4}\;_3F_2(1/2,1/2,1/2;1,3/2;x^2).\eqno(21)$$
Therefore, by setting $F={\bf K}$ in (9), we obtain
$$\int_0^1\frac{udu}{\sqrt{1-u^2}}{\bf K}^2\left(\sqrt{\frac{1-\sqrt{1-x^2u^2}}{2}}\right)=\frac{\pi^2}{4}\;_3F_2(1/2,1/2,1/2;1,3/2;x^2)\eqno(22)$$
and for $x=1$, after some simplification,
$$\int_0^{1/\sqrt{2}}k{\bf K}^2(k)dk=\frac{1}{4}\pi{\bf G}\eqno(23)$$
where ${\bf G}$ is Catalan's constant. Both (22) and (23) appear to be new and there is evidence that (23) is the only analytically tractable moment of ${\bf K}^2$ over a sub-unit interval[8].

In the same vein, let us set $F(t)=1/\sqrt{1-t^2}$ in (11) with $x=1$. This results in
$$\int_0^1\int_0^1\int_0^1\frac{ududvdw}{\sqrt{(1-u^2)(1-v^2)(1-w^2)(1-u^2v^2w^2)}}=\pi{\bf G}.\eqno(24)$$
An obvious application of (11) and its higher dimensional generalizations is to so-called Watson integrals [9], which are integrals over a polytope of a ratio of trigonometric polynomials. Thus, (11) gives immediately
$$\int_0^{\pi/2}\int_0^{\pi/2}\int_0^{\pi/2}\frac{\sin\beta \; d\beta d\phi d\theta}{1-x\sin\beta\sin\phi\sin\theta}=$$
$$\frac{\pi}{4x}\left[ArcCos^2(x)-2\pi ArcCos(x)+\frac{3\pi^2}{4}\right]\eqno(25)$$

An alternative form of (16) is
$$\int_0^1{\bf K}(u)f(u)du=\int_0^{\pi/2}\int_0^{\pi/2}\frac{d\theta d\phi}{1+\sin\theta\sin\phi}f\left(\frac{1-\sin\theta\sin\phi}{1+\sin\theta\sin\phi}\right).\eqno(26)$$
Reference [7] contains over 100 integrals having the form of the left hand side of (26) allowing many, apparently new, double trigonometric integrals to be found. For example,
$$\int_0^{\pi/2}\int_0^{\pi/2}\frac{d\theta d\phi}{\sqrt{\sin\theta\sin\phi(1+\sin\theta\sin\phi)}}=4k{\bf K}(k){\bf K}'(k),\eqno(27)$$
where $k=\sqrt{2}-1$.
With judicious selection of rational functions $F$ many striking, and potentially useful, results can be worked out in this area, which will be the subject of a future report.

\section*{Acknowledgements}\vskip .1in

The author thanks Prof. L.M. Nieto and the Physics Department of the University of Valladolid for their hospitality while this work was carried out.

\newpage
\centerline{\bf References}\vskip .2in
\noindent
[1] I.S. Gradshteyn and I.M. Ryzhik, {\it Table of Integrals, Series and Products}[ Academic Press, NY 1963]

\noindent
[2] {\it Tables of Integral Transforms I} Ed. A. Erd\'elyi,[McGraw-Hill Book Company, New York, 1954] Eq 4.3(12)

\noindent
[3] L.H. Loomis, {\it An Introduction to Abstract Harmonic Analysis}, [D.  Van Nostrand, Inc. New York, 1953]

\noindent
[4]  E.T. Whittaker and G.N. Watson, {\it A course in Modern Analysis},{Cambridge Univ. Press, 1927] Chap.11.

\noindent
[5]  David H. Bailey et al, J. Phys.{\bf A41}, 205203 (2008)

\noindent
[6]  D.Borwein et al., {\it Moments of Ramanujan's Generalized Elliptic Integrals} (Unpublished)

\noindent
[7] M.L. Glasser,  J. Res. of the NBS {\bf 80B}, 207 (1976).

\noindent
[8]  Dr. James Wan (Private communication). 

\noindent
[9] A.J. Guttmann, J.Phys.{\bf A42}, 232001(2009).

\end{document}